\documentclass[twocolumn,showpacs,preprintnumbers,amsmath,amssymb]{revtex4}
\usepackage{graphicx}
\usepackage{dcolumn}
\usepackage{bm}

\begin{document}

\title{Wave equations for determining energy-level gaps of
quantum systems}
\author{Zeqian Chen}
\email{chenzq@mail.cnu.edu.cn}
\affiliation{%
Research Center of Mathematical Physics, School of Mathematical
Sciences, Capital Normal University, 105 North Road, Xi-San-Huan,
Beijing}
\affiliation{%
Wuhan Institute of Physics and Mathematics, Chinese Academy of
Sciences, 30 West District, Xiao-Hong-Shan, P.O.Box 71010, Wuhan}

\date{\today}

\begin{abstract}
An differential equation for wave functions is proposed, which is
equivalent to Schr\"{o}dinger's wave equation and can be used to
determine energy-level gaps of quantum systems. Contrary to
Schr\"{o}dinger's wave equation, this equation is on `bipartite'
wave functions. It is shown that those `bipartite' wave functions
satisfy all the basic properties of Schr\"{o}dinger's wave
functions. Further, it is argued that `bipartite' wave functions can
present a mathematical expression of wave-particle duality. This
provides an alternative approach to the mathematical formalism of
quantum mechanics.
\end{abstract}

\maketitle

In the most general form, Heisenberg's equation \cite{H} and
Schr\"{o}dinger's equation \cite{S} can be written as
follows\begin{equation}i\hbar \frac{\partial \hat{O} (t)}{\partial
t } = \left [ \hat{O} (t), \hat{H} \right
],\end{equation}and\begin{equation}i\hbar \frac{\partial | \psi
(t) \rangle}{\partial t} = \hat{H} | \psi (t)
\rangle,\end{equation}respectively, where $\hat{H}$ is the
Hamiltonian of the system. As is well known, these two forms for
the equations of motion of quantum mechanics are equivalent. Of
these, the Schr\"{o}dinger form seems to be the more useful one
for practical problems, as it provides differential equations for
wave functions, while Heisenberg's equation involves as unknowns
the operators forming the representative of the dynamical
variable, which are far more numerous and therefore more difficult
to evaluate than the Schr\"{o}dinger unknowns.

On the other hand, determining energy levels of various dynamic
systems is an important task in quantum mechanics, for this solving
Schr\"{o}dinger's wave equation is a usual way. Recently, Fan and Li
\cite{FL} showed that Heisenberg's equation can also be used to
deduce the energy level of some systems. By introducing the
conception of invariant `eigen-operator', they derive energy-level
gap formulas for some dynamic Hamiltonians. However, their
`invariant eigen-operator' equation involves operators as unknowns,
as similar to Heisenberg's equation, and hence is also difficult to
evaluate in general.

In this article we propose an differential equation for wave
functions, which can be used to determine energy-level gaps of
quantum systems and is mathematically equivalent to
Schr\"{o}dinger's wave equation, that is, they can be solved from
one another. Contrary to Schr\"{o}dinger's wave equation, this
equation is on `bipartite' wave functions. It is shown that those
`bipartite' wave functions satisfy all the basic properties of
Schr\"{o}dinger's wave functions. In particular, it is argued that
`bipartite' wave functions can present a mathematical expression of
wave-particle. This provides an alternative approach to the
mathematical formalism of quantum mechanics.

For convenience, we deal with the quantum system of a single
particle. Note that the Hamiltonian for a single particle in an
external field is\begin{equation}\hat{H}(\vec{x}) = -
\frac{\hbar^2}{ 2 m} \nabla^2_{\vec{x}} + U(\vec{x}
),\end{equation}where $\nabla^2_{\vec{x}} = \partial^2/\partial
x^2_1 +
\partial^2/\partial x^2_2 + \partial^2/\partial x^2_3,$
$U(\vec{x})$ is the potential energy of the particle in the
external field, and $\vec{x} = (x_1, x_2, x_3) \in \mathbb{R}^3.$
Then, Schr\"{o}dinger's wave equation for a single particle in an
external field is\begin{equation}i\hbar \frac{\partial \psi
(\vec{x}, t) }{\partial t} = \hat{H}(\vec{x}) \psi (\vec{x}, t) =
- \frac{\hbar^2}{ 2 m} \nabla^2_{\vec{x}} \psi (\vec{x}, t) +
U(\vec{x} ) \psi (\vec{x}, t).\end{equation} On the other hand,
let $\psi (\vec{x}, t)$ and $\varphi (\vec{x}, t)$ both satisfy
Eq.(4). Then we have$$\begin{array}{lcl}i\hbar \frac{\partial (
\psi (\vec{x}, t) \varphi^* (\vec{y}, t)) }{\partial t} & = &
i\hbar \frac{\partial \psi (\vec{x}, t) }{\partial t} \varphi^*
(\vec{y}, t) + i\hbar \frac{\partial \varphi^* (\vec{y}, t)
}{\partial t} \psi (\vec{x}, t)\\[0.4mm]
& = & \left [ \hat{H}(\vec{x}) \psi (\vec{x}, t) \right ]
\varphi^* (\vec{y}, t)\\[0.4mm]
&~&~~ - \left [ \hat{H}(\vec{y}) \varphi
(\vec{y}, t) \right ]^* \psi (\vec{x}, t)\\[0.4mm]
& = & \left ( \hat{H}(\vec{x}) - \hat{H}(\vec{y}) \right ) (\psi
(\vec{x}, t) \varphi^* (\vec{y}, t)).
\end{array}$$This leads to the following wave equation
\begin{equation}i\hbar \frac{\partial \Psi
(\vec{x}, \vec{y}; t) }{\partial t} = \left (\hat{H}(\vec{x}) -
\hat{H}(\vec{y}) \right ) \Psi (\vec{x}, \vec{y};
t),\end{equation}where $\Psi (\vec{x}, \vec{y}; t ) \in
L^2_{\vec{x}, \vec{y}}.$ Contrary to Schr\"{o}dinger's wave
equation Eq.(4) for `one-partite' wave functions $\psi (\vec{x})
\in L^2_{\vec{x}},$ the wave equation Eq.(5) is an differential
equation for `bipartite' wave functions $\Psi (\vec{x}, \vec{y}),$
which, replacing $\hat{H}(\vec{x}) + \hat{H}(\vec{y})$ by
$\hat{H}(\vec{x}) - \hat{H}(\vec{y}),$ is also different from
Schr\"{o}dinger's wave equation for two particles.

Since$$\frac{\partial \left | \Psi (\vec{x}, \vec{y}; t) \right
|^2}{\partial t} = 2 i \mathrm{Re} \left [ \Psi^* (\vec{x}, \vec{y};
t) \frac{\partial \Psi (\vec{x}, \vec{y}; t) }{\partial t} \right
],$$it is concluded from Eq.(5) that\begin{equation} \frac{\partial
}{\partial t} \int \left | \Psi (\vec{x}, \vec{y}; t) \right |^2
d^3\vec{x} d^3 \vec{y} = 0.\end{equation}This implies that Eq.(5)
preserves the probability density $\left | \Psi (\vec{x}, \vec{y};
t) \right |^2 d^3\vec{x} d^3 \vec{y}$ with respect to time and means
that, if this wave function $\Psi$ is given at some instant, its
behavior at all subsequent instants is determined.

By Schmidt's decomposition theorem \cite{Schmidt}, for every $\Psi
(\vec{x}, \vec{y}) \in L^2_{\vec{x}, \vec{y}}$ there exist two
orthogonal sets $\{\psi_n \}$ and $\{\varphi_n \}$ in
$L^2_{\vec{x}}$ and $L^2_{\vec{y}}$ respectively, and a sequence of
positive numbers $\{ \mu_n \}$ satisfying $\sum_n \mu^2_n  < \infty$
so that\begin{equation}\Psi (\vec{x}, \vec{y}) = \sum_n \mu_n \psi_n
(\vec{x}) \varphi^*_n (\vec{y}).\end{equation}Then, it is easy to
check that$$\Psi (\vec{x}, \vec{y}; t) = \sum_n \mu_n \psi_n
(\vec{x}, t) \varphi^*_n (\vec{y}, t)$$satisfies Eq.(5) with $\Psi
(\vec{x}, \vec{y}; 0) = \Psi (\vec{x}, \vec{y}),$ where both $\psi_n
(\vec{x}, t)$ and $\varphi_n (\vec{y}, t)$ satisfy Eq.(4) with
$\psi_n (\vec{x}, 0) = \psi_n (\vec{x})$ and $\varphi_n (\vec{y}, 0)
= \varphi_n (\vec{y}),$ respectively. Hence, the wave equation
Eq.(5) can be solved mathematically from Schr\"{o}dinger's wave
equation.

Given $\psi \in L^2_{\vec{x}},$ for every $t \geq 0$ define
operators $\varrho_t$ on $L^2_{\vec{x}}$
by\begin{equation}(\varrho_t \varphi ) (\vec{x}) = \int \Psi
(\vec{x}, \vec{y}; t) \varphi (\vec{y}) d^3
\vec{y},\end{equation}where $\Psi (\vec{x}, \vec{y}; t)$ is the
solution of Eq.(5) with $\Psi (\vec{x}, \vec{y}; 0) = \psi
(\vec{x}) \psi^* (\vec{y}).$ It is easy to check
that\begin{equation}i \frac{\partial \varrho_t}{\partial t } =
\left [ H, \varrho_t \right ],~~\varrho_0 = | \psi \rangle \langle
\psi |.\end{equation}This is just Schr\"{o}dinger's equation in
the form of density operators. Hence, Schr\"{o}dinger's wave
equation is a special case of the wave equation Eq.(5) with
initial values of product form $\Psi (\vec{x}, \vec{y}; 0) = \psi
(\vec{x}) \psi^* (\vec{y}).$ Therefore, the wave equation Eq.(5)
is mathematically equivalent to Schr\"{o}dinger's wave equation.

In the sequel, we consider the problem of stationary states. Let
$\psi_n$ be the eigenfuncions of the Hamiltonian operator
$\hat{H},$ i.e., which satisfy the equation\begin{equation}
\hat{H}(\vec{x}) \psi_n (\vec{x}) = E_n \psi_n
(\vec{x}),\end{equation}where $E_n$ are the eigenvalues of
$\hat{H}.$ Correspondingly, the wave equation
Eq.(5)$$\begin{array}{lcl}i\hbar \frac{\partial \Psi (\vec{x},
\vec{y}; t) }{\partial t} &=& \left (\hat{H}(\vec{x}) -
\hat{H}(\vec{y}) \right ) \Psi (\vec{x}, \vec{y}; t)\\& = &(E_n -
E_m ) \Psi (\vec{x}, \vec{y}; t)\end{array}$$with $\Psi (\vec{x},
\vec{y}; 0) = \psi_n (\vec{x}) \psi^*_m (\vec{y}),$ can be
integrated at once with respect to time and
gives\begin{equation}\Psi (\vec{x}, \vec{y}; t) =
e^{-i\frac{1}{\hbar} (E_n - E_m) t} \psi_n (\vec{x}) \psi^*_m
(\vec{y}).\end{equation}Since $\{ \psi_n (\vec{x}) \}$ is a
complete orthogonal set in $L^2_{\vec{x}},$ it is concluded that
$\{ \psi_n (\vec{x}) \psi^*_m (\vec{y}) \}$ is a complete
orthogonal set in $L^2_{\vec{x}, \vec{y}}.$ Then, for every $\Psi
(\vec{x}, \vec{y}) \in L^2_{\vec{x}, \vec{y}}$ there exists a
unique set of numbers $\{ c_{n,m}\}$ satisfying $\sum_{n,m} |
c_{n,m} |^2 < \infty$ so that
\begin{equation}\Psi (\vec{x}, \vec{y}) = \sum_{n,m} c_{n,m} \psi_n (\vec{x})
\psi^*_m (\vec{y}).\end{equation}Hence, for $\Psi (\vec{x},
\vec{y}; 0) = \sum_{n,m} c_{n,m} \psi_n (\vec{x}) \psi^*_m
(\vec{y})$ we have that\begin{equation}\Psi (\vec{x}, \vec{y}; t)
= \sum_{n,m} c_{n,m} e^{-i\frac{1}{\hbar} (E_n - E_m) t} \psi_n
(\vec{x}) \psi^*_m (\vec{y})\end{equation}for $t \geq 0.$ Now, if
$\Psi (\vec{x}, \vec{y}) \in L^2_{\vec{x}, \vec{y}}$ is an
eigenfuncion of the operator $\hat{H}(\vec{x})- \hat{H}(\vec{y}),$
i.e., which satisfies the equation\begin{equation} \left (
\hat{H}(\vec{x})- \hat{H}(\vec{y}) \right ) \Psi (\vec{x},
\vec{y}) = \lambda \Psi (\vec{x}, \vec{y}),\end{equation}where
$\lambda$ is an associated eigenvalue, then $\Psi (\vec{x},
\vec{y}; t) = e^{-i\frac{1}{\hbar} \lambda t} \Psi (\vec{x},
\vec{y})$ satisfies Eq.(5) and consequently, it is concluded from
Eq.(13) that $\lambda = E_n - E_m$ is an energy-level gap of the
system. Thus, the wave equation Eq.(5) can be used to determine
energy-level gaps of the system.

It is well known that the basis of the mathematical formalism of
quantum mechanics lies in the proposition that the state of a
system can be described by a definite Schr\"{o}dinger's wave
function of coordinates \cite{vN,Dirac}. The square of the modulus
of this function determines the probability distribution of the
values of the coordinates \cite{Born}. Since the wave equation
Eq.(5) is mathematically equivalent to Schr\"{o}dinger's wave
equation, it seems that the state of a quantum system also can be
described by a definite `bipartite' wave function of Eq.(5), of
which the physical meaning is that the `bipartite' wave functions
of stationary states determine energy-level gaps of the system.

In fact, we can make the general assumption that if the measurement
of an observable $\hat{O}$ for the system in the `bipartite' state
corresponding to $\Psi$ is made a large number of times, the average
of all the results obtained will be\begin{equation} \langle \hat{O}
\rangle_{\Psi} = \mathrm{Tr} \left [ \varrho^{\dagger}_{\Psi}
\hat{O} \varrho_{\Psi} \right ],\end{equation}where $\varrho_{\Psi}$
is an operator on $L^2$ associated with $\Psi$ defined by
$(\varrho_{\Psi} \varphi ) (\vec{x}) = \int \Psi (\vec{x}, \vec{y})
\varphi (\vec{y}) d^3 \vec{y}$ for every $\varphi \in L^2,$ provided
$\Psi$ is normalized. That is, the expectation value of an
observable $\hat{O}$ in the `bipartite' state corresponding to
$\Psi$ is determined by Eq.(15). It is easy to check that if $\Psi
(\vec{x}, \vec{y}) = \psi (\vec{x}) \psi^* (\vec{y}),$
then\begin{equation} \langle \hat{O} \rangle_{\Psi} = \langle \psi |
\hat{O} | \psi \rangle.\end{equation}This concludes that our
expression Eq.(15) agrees with the interpretation of
Schr\"{o}dinger's wave functions for calculating expectation values
of any chosen observable.

Moreover, `bipartite' wave functions can present a mathematical
expression of wave-particle duality. Let us discuss the double-slit
experiment \cite{Feynman}. Let $\phi_1$ and $\phi_2$ be two
Schr\"{o}dinger's wave functions of the particle arrival through
slit 1 and slit 2, respectively. Then, the associated `bipartite'
wave function of the particle arrival through both slit 1 and slit 2
can be either\begin{equation}\Psi_{\mathrm{W}} (\vec{x}, \vec{y}) =
\left ( \phi_1 (\vec{x}) + \phi_2 (\vec{x}) \right ) \left (
\phi^*_1 (\vec{y}) + \phi^*_2 (\vec{y}) \right
),\end{equation}or\begin{equation}\Psi_{\mathrm{P}} (\vec{x},
\vec{y}) = \phi_1 (\vec{x}) \phi^*_1 (\vec{y}) + \phi_2 (\vec{x})
\phi^*_2 (\vec{y}).\end{equation}A single particle described by
$\Psi_{\mathrm{W}}$ behaves like waves, while by $\Psi_{\mathrm{P}}$
like particles. This is so because for position, by (15) we
have$$\langle \hat{x}\rangle_{\Psi_{\mathrm{W}}} \propto |\phi_1
(\vec{x}) + \phi_2 (\vec{x})|^2, \langle
\hat{x}\rangle_{\Psi_{\mathrm{P}}} \propto |\phi_1 (\vec{x})|^2 + |
\phi_2 (\vec{x})|^2$$respectively. On the other hand,
$\Psi_{\mathrm{P}}$ is a `bipartite' entangled state \cite{EPR},
which means that a single particle can entangle with itself
\cite{Enk}, as similar to the fact that each photon can interfere
with itself, as shown in Ref. \cite{Dirac}. Then, it is concluded
that a single particle behaves like waves when it interfere with
itself, while like particle when entangle with itself. Thus,
wave-particle duality is just the complementarity of interference
and entanglement for a single particle. A more detail on this issue
will be given in the future. Since entanglement plays a crucial role
in quantum communication, cryptograph, and computation \cite{B-G-N},
we may expect that the entanglement of a {\it single} particle will
play an important role in quantum information \cite{Babichev}.

We would like to mention that Eq.(5) have been presented by Landau
and Lifshitz \cite{LL} giving the change in the density matrix with
time, similar to the Schr\"{o}dinger's wave equation. However, we
regard Eq.(5) as a wave equation but not a equation for density
functions. This is the key point which is distinct from \cite{LL}.
As shown above, Eq.(5) is a suitable form for motion of quantum
mechanics as a `bipartite' wave equation.

In summary, we present an differential equation for wave functions,
which is equivalent to Schr\"{o}dinger's wave equation and can be
used to determine energy-level gaps of the system. Contrary to
Schr\"{o}dinger's wave equation, this equation is on `bipartite'
wave functions. It is shown that those `bipartite' wave functions
satisfy all the basic properties of Schr\"{o}dinger's wave
functions. Further, it is argued that `bipartite' wave functions can
present a mathematical expression of wave-particle duality. Our
results shed considerable light on the mathematical basis of quantum
mechanics.

This work was supported by the National Natural Science Foundation
of China under Grant No.10571176, the National Basic Research
Programme of China under Grant No.2001CB309309, and also funds
from Chinese Academy of Sciences.


\end{document}